\documentclass[aps,nofootinbib,raggedbottom]{revtex4-1}
\pdfoutput=1
\usepackage{graphicx,subfigure}
\usepackage{amsmath,amssymb}
\usepackage{xcolor}
\usepackage{mathrsfs}  
\usepackage{ulem}

\def\beq{\begin{equation}}
	\def\eeq{\end{equation}}
\def\beqr{\begin{eqnarray}}
	\def\eeqr{\end{eqnarray}}
\def\bdpm{\begin{displaymath}}
	\def\edpm{\end{displaymath}}
\def\half{\frac{1}{2}}
\definecolor{lgray}{gray}{0.6}

\newcommand{\nnb}{\nonumber}

\newcommand{\tth}{t_\theta}
\newcommand{\cth}{c_\theta}
\newcommand{\sth}{s_\theta}

\newcommand{\Br}{\mathrm{Br}}
\newcommand{\BR}{\mathcal{B}}

\begin{document}

\title{
Mono-Higgs signature in a singlet fermionic dark matter model
}

\author{Yeong Gyun Kim}
\email{ygkim@gnue.ac.kr}

\affiliation{ 
	Department of Science Education, 
	Gwangju National University of Education, Gwangju 61204, Korea
}

\author{Kang Young Lee}
\email{kylee.phys@gnu.ac.kr; corresponding author}

\affiliation{ 
	Department of Physics Education \& RINS,
	Gyeongsang National University, Jinju 52828, Korea
}

\author{Soo-hyeon Nam}
\email{glvnsh@gmail.com; corresponding author}

\affiliation{ 
	The Institute of Basic Science, 
	Korea University, Seoul 02841, Korea
}

\date{\today}

\begin{abstract}
	
We investigate mono-Higgs production as a probe of singlet fermionic dark matter (SFDM) at the LHC. 
In this framework, a Standard Model (SM) gauge-singlet Dirac fermion serves as the dark matter candidate, 
interacting with the visible sector through a real scalar mediator that mixes with the SM Higgs boson. 
Focusing on light dark matter with $M_\psi \in [0.1,10]$ GeV,
we analyze the viable parameter space under constraints from relic density, Higgs decay properties, 
invisible decay bounds, rare $B$-meson decays, and direct detection experiments.
We compute the mono-Higgs production cross sections at
$\sqrt{s}=13~\mathrm{TeV}$ and compare approximate detector-level yields
with the ATLAS $b\bar b$ search, supplemented by an inclusive upper bound
for the CMS four-lepton channel. 
For the two benchmarks, the signal is dominated by non-resonant Higgs-pair production 
followed by an invisible Higgs decay. The resulting signal rate depends on both the production cross sections 
and the physical invisible branching fractions. The inclusive rates of $3.33$ and $1.43$~fb
at loop-induced leading order provide quantitative signal targets before
the visible Higgs decay and analysis cuts. 
For representative benchmark points satisfying the constraints
considered, these comparisons indicate limited sensitivity in the analyzed channels.
Nevertheless, the mono-Higgs channel provides a complementary probe of
Higgs-portal dark matter scenarios, particularly in the low-mass mediator regime.

\end{abstract}

\maketitle

\section{Introduction}

The existence of dark matter (DM) is strongly supported by numerous astrophysical and cosmological observations, 
such as galaxy rotation curves, the Bullet Cluster, gravitational lensing, 
and anisotropies in the cosmic microwave background (CMB)
\cite{Bertone:2016nfn,Bauer:2017qwy,Mambrini:2021cwd,Marsh:2024ury,Cirelli:2024ssz}.
We now seek to understand whether DM particles interact with Standard Model (SM) particles and, 
if so, what signatures such interactions may produce.
A massive DM particle that interacts with SM particles with a strength comparable to that of the weak interaction,
known as a weakly interacting massive particle (WIMP), is one of the most popular DM candidates.

If there exist non-gravitational interactions between DM particles and the SM matter,
it may be possible to search for DM at colliders.
A possible collider signature of WIMPs is the production of a single visible SM particle 
accompanied by large missing transverse energy ($E^{\textrm{miss}}_{\textrm{T}}$).
Since DM particles escape the detector without interacting, 
they leave behind an imbalance in the transverse momentum.
Such signatures occur in many DM scenarios, although their
rates and kinematic distributions depend on the underlying model.
Collider studies have focused on event topologies where a single SM particle is produced 
in association with DM particles, leading to so-called mono-$X$ signatures.
To date, a variety of mono-$X$ signatures with $E^{\textrm{miss}}_{\textrm{T}}$ have been explored, 
such as mono-jet \cite{ATLAS:2021kxv,CMS:2017zts,CDF:2012tfi},
mono-photon \cite{ATLAS:2020uiq,CMS:2017qyo,CDF:2008njt},
and mono-$Z/W$ boson \cite{ATLAS:2018nda,CMS:2017jdm,ATLAS:2017nyv} events.

In this work, we consider the singlet fermionic DM (SFDM) model mediated by an additional real scalar field, 
as studied in Refs. \cite{Kim:2006af,Kim:2008pp,Kim:2016csm}.
The scalar field mixes with the SM Higgs field, resulting in two neutral scalar bosons ($h_1$, $h_2$) in the model.
The Higgs field serves as the only portal connecting SM matter to the dark sector, 
and such models are commonly referred to as Higgs portal models.
Since the Higgs boson couples directly to the dark sector in this setup, 
mono-Higgs events can probe the Higgs-portal interaction.
Moreover, the mono-Higgs production via initial state radiation is highly suppressed 
due to the small Yukawa couplings of light quarks.
As a result, observation of these events would provide a direct probe 
of the interaction between the dark sector and the SM Higgs.
In this context, the ATLAS and CMS collaborations have recently investigated various decay modes 
of the Higgs boson to search for mono-Higgs production.
Among these, the decay of the Higgs boson $h_1$ into a $b \bar{b}$ pair has the largest branching ratio
\cite{ATLAS:2021shl,ATLAS:2017uis,ATLAS:2016cfj,ATLAS:2015ptl,CMS:2018zjv,CMS:2017prz},
while the decays into $\gamma \gamma$ and $\tau^- \tau^+$ provide cleaner signals
\cite{CMS:2019ykj,ATLAS:2021jbf,ATLAS:2017pzz,ATLAS:2015wpv,CMS:2018nlv,ATLAS:2023ild}.
The $(h\to b\bar b)+(h\to{\rm inv.})$ signature through Higgs-pair
production was studied in Ref.~\cite{Banerjee:2016nzb}, both for SM-like
non-resonant production and for production enhanced by a heavy di-Higgs
resonance. Our new results concern light SFDM benchmarks whose invisible
branching fractions and mono-Higgs rates are correlated with the thermal
relic abundance, Higgs properties, mass-dependent invisible-meson decay
bounds and low-mass direct detection.
We include low-energy hadronic annihilation channels and compare the
benchmark relic densities under different hadronic prescriptions.
For these benchmarks, $h_2\to h_1h_1$ is kinematically closed ($M_2<2M_1$),
and the signal is dominated by non-resonant $h_1h_1$ production with an invisible decay.
We quantify their yields and the limitations of the ATLAS $b\bar b$
and CMS four-lepton channels.

In this SFDM model, we focus on light DM with $M_\psi \in [0.1,10]$ GeV.
This range includes the sub-GeV region, where recoil-energy thresholds substantially reduce
the sensitivity of nuclear-recoil searches, and the few-GeV region, where direct detection
provides an important complementary test.
The precise measurement of the relic density also places strong constraints on the DM properties.
In our analysis, we find that the viable parameter space is mostly limited to the resonance region, 
with the mass of the additional scalar $h_2$ being nearly twice the DM mass.
For the selected benchmarks, $h_2$ decays predominantly
into a DM pair, allowing both direct and cascade contributions to the
mono-Higgs signature.
Furthermore, the measured total width and invisible branching ratio of the Higgs boson 
impose stringent constraints on the model parameter space, 
since the SM-like Higgs boson can have a sizable branching ratio into a pair of $h_2$ bosons 
when $h_2$ is sufficiently light.
In addition, rare $B$-meson decays induced by $h_2$ provide complementary constraints on the model, 
restricting the scalar mixing angle, $\theta$.
A similar analysis has been performed in a model with a pseudoscalar mediator \cite{Ghorbani:2016edw}. 
However, the study focused on the parameter region where the fermionic DM is relatively heavy, 
leading to phenomenology different from that considered in our analysis.
Pseudoscalar fermion interactions can suppress nonrelativistic DM scattering by powers
of the momentum transfer~\cite{GAMBIT:2019HiggsPortal}.
The purely scalar Yukawa interaction retained here has no such leading-order suppression.
This makes the low-mass region particularly relevant, while invisible meson decays impose
complementary restrictions on the light scalar portal.

The recent analyses by ATLAS~\cite{ATLAS:2021shl} and CMS~\cite{CMS:2019ykj} investigated mono-Higgs production 
at a center-of-mass (CM) energy of 13 TeV, with integrated luminosities of 139 fb$^{-1}$ and 35.9 fb$^{-1}$, respectively, 
and set exclusion limits on two benchmark models.
Taking into account the constraints on the model parameters,
we estimate SFDM yields with a partial implementation of the ATLAS selection
and bound the CMS four-lepton yield before acceptance losses.
We explicitly distinguish these estimates from the full experimental selections.
This paper is organized as follows.
In Sec. II, we describe the model and define the relevant parameters.
In Sec. III, we discuss the constraints used to select the model parameters.
In Sec. IV, we present the mono-Higgs cross sections,
the detector-level comparisons, and their limitations.
Finally, in Sec. V, we summarize our conclusions.

\section{The Model}

In this paper, we adopt a DM model consisting of a Dirac fermion field $\psi$ and a real scalar field $S$,
both of which are SM gauge singlets.
The fermion $\psi$ serves as the DM candidate, 
while the scalar $S$ plays the role of a mediator between the dark and visible sectors.
As the model has been thoroughly discussed in Refs. \cite{Kim:2006af,Kim:2008pp,Kim:2016csm}, 
we summarize only the key aspects relevant to our analysis.
The dark sector Lagrangian with the renormalizable interactions 
is then given by
\beq \label{eq:DM_Lagrangian}
\mathcal{L}_{DM} = \bar{\psi}\left(i\partial\!\!\!/ 
                  - M_{\psi_0}\right)\psi 
                 + \half \left(\partial_{\mu} S\right)\left(\partial^{\mu} 
		 S\right)  - g_S \bar{\psi}\psi S - V_{S}(S,H), 
\eeq
where the Higgs portal potential is
\beq \label{eq:VS_potential}
V_{S}(S,H) = \half M_S^2 S^2 + \lambda_1 H^{\dagger} H S 
             + \lambda_2 H^{\dagger} H S^2 
             + \frac{\lambda_3}{3!}S^3 + \frac{\lambda_4}{4!}S^4 .
\eeq
Note that the Higgs quadratic part $H^{\dagger}H$ is 
the only portal to the singlet sector. 

After electroweak symmetry breaking, 
the neutral component of the SM Higgs and the singlet scalar 
develop nonzero vacuum expectation values (VEVs),  
$\langle H^0 \rangle=v_h/\sqrt{2}$ and 
$\langle S \rangle = v_s$, respectively.
By minimizing the full scalar potentials $V_S + V_{SM}$, where
\begin{eqnarray}
	V_{SM} = -\mu^2 H^{\dagger} H + \lambda_0 (H^{\dagger} H)^2,
\end{eqnarray}
the scalar mass parameters $M_S^2$ and $\mu^2$ can be expressed in terms of the scalar VEVs as follows \cite{PMS}:
\beq
M_S^2 = - \left(\frac{\lambda_1}{2 v_s} 
          + \lambda_2 \right)v_h^2 - \left(\frac{\lambda_3}{2v_s} 
	  + \frac{\lambda_4}{6}\right) v_s^2 ,
\quad
\mu^2 = {\lambda}_0 v_h^2 + (\lambda_1 + \lambda_2 v_s)v_s .
\eeq
The neutral scalar fields $h$ and $s$ defined by 
$H^0=(v_h+h)/\sqrt{2}$ and $S=v_s+s$ are mixed to yield 
the mass matrix given by
\beq 
	\mu_{h}^2 = 2 {\lambda}_0 v_h^2 , 
\quad
	\mu_{s}^2 = - \frac{\lambda_1 v_h^2}{2 v_s} 
	        + \frac{\lambda_3}{2} v_s + \frac{\lambda_4}{3} v_s^2 , 
\quad
	\mu_{hs}^2 =  (\lambda_1 + 2 \lambda_2 v_s) v_h.
\eeq
The corresponding scalar mass eigenstates $h_1$ and $h_2$ 
are admixtures of $h$ and $s$,
\beq
\left( \begin{array}{c} h_1 \\[1pt] h_2 \end{array} \right) =
\left( \begin{array}{cc} \cos \theta &\ \sin \theta \\[1pt]
	- \sin \theta &\ \cos \theta \end{array} \right)
\left( \begin{array}{c} h \\[1pt] s \end{array} \right) ,
\eeq
where the mixing angle $\theta$ is given by
\begin{eqnarray}
	\tan\theta = - \frac{\mu_h^2 - \mu_s^2 - \sqrt{(\mu_h^2-\mu_s^2)^2 + 4\mu_{hs}^4 }}{2\mu_{hs}^2}.
\end{eqnarray}
After the mass matrix is diagonalized, 
we obtain the physical masses of the two scalar bosons $h_1$ and $h_2$ as follows:
\beq \label{eq:scalar_mass}
M_1^2 = \frac{\mu_h^2 - \mu_s^2\tth^2}{1-\tth^2}, \quad
M_2^2 = \frac{\mu_s^2 - \mu_h^2\tth^2}{1-\tth^2},
\eeq
 where $t_\theta \equiv \tan\theta$.
We assume that $M_1$ corresponds to the observed SM-like Higgs boson mass in what follows.

The model contains eight independent parameters relevant for dark matter (DM) phenomenology.
The singlet fermion mass is given by $M_\psi = M_{\psi_0} + g_S v_s$, 
where $M_{\psi_0} $ is a free parameter 
and the Yukawa coupling $g_S$ controls the interaction strength between the Dirac fermion $\psi$ 
and the singlet component of the scalar sector. 
We treat $M_\psi$ as an independent parameter.
The six parameters in the scalar potential, $\lambda_0, \lambda_1, \lambda_2, \lambda_3, \lambda_4$, and $v_s$,
determine the scalar masses $M_{1,2}$, the mixing angle $\theta$, 
and the self-interactions of the two physical scalars $h_1$ and $h_2$. 
With the SM-like Higgs boson mass fixed at $M_1 =125$ GeV, 
we are left with seven independent new physics parameters. 
In the next section, we apply Higgs, flavor, relic-density
and direct-detection constraints to the model parameters.

\section{Constraints}
\label{sec:constraints}

In this section, we examine the phenomenological constraints on the model 
from both collider experiments and dark matter (DM) observations.
Due to the Higgs-portal terms in Eq.~(\ref{eq:VS_potential}), 
the electroweak interaction of the Higgs boson can be significantly modified.
Depending on their masses, 
the scalar states $h_1$ and $h_2$ may decay into one another. 
To account for the Higgs decay width constraint, we derive the cubic Higgs self-coupling coefficients $c_{ijk}$ 
relevant for $h_i h_j h_k$ interactions as follows \cite{Kim:2016csm}: 
\beqr
c_{111}
&=& 6 \lambda_0 v_h \cth^3 + \left( 3 \lambda_1 + 6 \lambda_2
v_s \right) \cth^2 \sth + 6 \lambda_2 v_h
\cth \sth^2 + (\lambda_3 + \lambda_4 v_s)
\sth^3 , \nnb\\
c_{112}
&=& -6 \lambda_0 v_h \cth^2 \sth + 2 \lambda_2 v_h
\left( 2 \cth^2 \sth - \sth^3\right) 
    + \left(\lambda_1 + 2 \lambda_2 v_s \right) \left( \cth^3 -
2 \cth \sth^2 \right) + \left( \lambda_3 + \lambda_4 v_s
\right) \cth \sth^2 , \nnb\\
c_{122}
&=& 6 \lambda_0 v_h \cth \sth^2 + 2 \lambda_2 v_h
\left( \cth^3 - 2 \cth \sth^2\right) 
     - \left(\lambda_1 + 2 \lambda_2 v_s \right) \left( 2 \cth^2
\sth - \sth^3 \right) + \left( \lambda_3 +
\lambda_4 v_s \right) \cth^2 \sth , \nnb\\
c_{222}
&=& - 6 \lambda_0 v_h \sth^3 
+ \left( 3 \lambda_1 + 6\lambda_2 v_s \right) \sth^2 \cth 
- 6 \lambda_2 v_h
\sth \cth^2  + (\lambda_3 + \lambda_4 v_s) \cth^3 ,
\label{eq:cijk}
\eeqr
where $s_\theta \equiv \sin\theta$ and $c_\theta \equiv \cos\theta$.

If $M_2 < M_1/2$, 
the decay $h_1 \to h_2 h_2$ is kinematically allowed via the triple coupling $c_{122}$.
The corresponding partial width is given by
\beq \label{eq:h1_h2h2_width}
\Gamma\left(h_1 \to h_2 h_2\right) 
     = \frac{|c_{122}|^2}{32\pi M_1}
     \left(1 - \frac{4M_2^2}{M_1^2}\right)^{1/2},  
\eeq
which increases the total decay width of $h_1$.
Therefore, the cubic coupling $c_{122}$ is strongly constrained 
by the current experimental measurement of the Higgs total decay width,
$\Gamma_\textrm{exp}= 3.7^{+1.9}_{-1.4}$ MeV 
\cite{ParticleDataGroup:2024cfk}.
In addition, when $M_\psi < M_1/2$, 
the SM-like Higgs boson $h_1$ can decay directly into a pair of DM fermions via scalar mixing.  
The corresponding partial decay width is given by
\beq \label{eq:inv_direct}
\Gamma(h_1 \to \psi\bar\psi) 
= \frac{g_S^2 M_1 \sin^2\theta}{8\pi} 
\left(1 - \frac{4M_\psi^2}{M_1^2} \right)^{3/2}.
\eeq
Moreover, if $M_\psi < M_2/2$, 
the cascade process $h_1 \to h_2 h_2 \to (\psi\bar{\psi})(\psi\bar{\psi})$ also contributes 
to the invisible decay width of $h_1$.
In this case, the total invisible branching ratio of $h_1$ becomes
\beq \label{eq:inv_BR}
\mathrm{BR}(h_1 \to \text{inv.}) =
\frac{
 \Gamma(h_1 \to \psi\bar{\psi}) +
 \Gamma(h_1 \to h_2 h_2)\,[\mathrm{BR}(h_2 \to \psi\bar\psi)]^2
}{
 \Gamma_{\text{SM}}\,\cth^2 +
 \Gamma(h_1 \to \psi\bar{\psi}) + \Gamma(h_1 \to h_2 h_2)
},
\eeq
where $\Gamma_{\text{SM}}=4.07~\text{MeV}$ is the SM total width at
$125$~GeV~\cite{LHCHiggsCrossSectionWorkingGroup:2013yr3}.
The factor $\cth^2$ accounts for the mixing suppression of the
SM decay modes. This normalization is also used in Sec.~IV.
Here the invisible branching fraction refers to the new direct and
cascade decays in Eq.~\eqref{eq:inv_BR}.
The cascade contribution vanishes when
$\mathrm{BR}(h_2\to\psi\bar\psi)\to0$.
We focus on $2M_\psi<M_2<M_1/2$ with predominantly invisible $h_2$ decays.
We adopt the ATLAS combined 95\% confidence-level limit
$\mathrm{BR}(h\to{\rm inv.})<0.107$~\cite{ATLAS:2023tkt},
obtained from $139~{\rm fb}^{-1}$ at $13$~TeV together with the
$7$ and $8$~TeV results.
Although the quoted limit assumes SM Higgs production, we use it as an approximation in the small-mixing regime considered here.
For the scalar mixing angle, we use the decay-mode-independent OPAL search
for $e^+e^- \to Zh_2$~\cite{OPAL:2002decay}.
Its 95\% confidence-level limit is expressed in terms of the production-rate ratio
$k=\sigma(e^+e^-\to Zh_2)/\sigma_{\rm SM}(e^+e^-\to Zh_{\rm SM};M_{h_{\rm SM}}=M_2)$.
In this model, $k=\sth^2$, so the OPAL result constrains the mixing angle
as a function of $M_2$, independently of the $h_2$ decay mode.

\begin{figure}[!t]
    \centering
    \includegraphics[width=12cm]{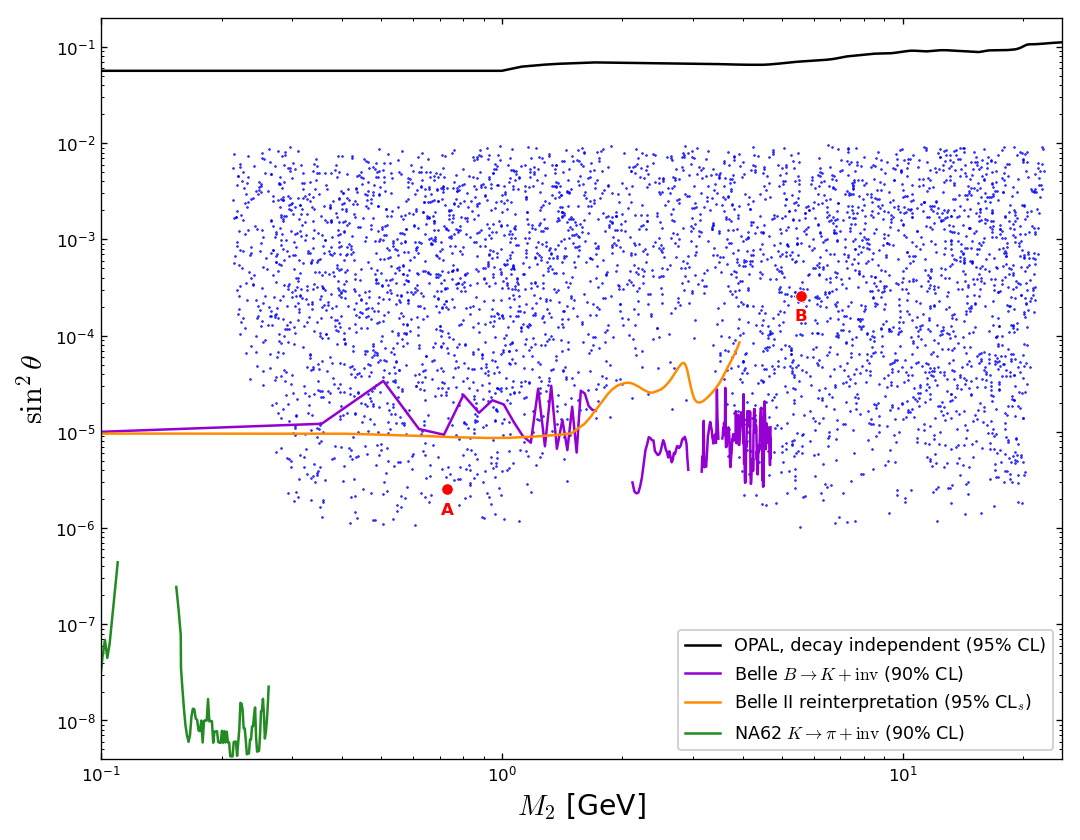}
    \caption{
    Experimental upper limits on $\sin^2\theta$ as a function of the scalar mass $M_2$:
    OPAL (black, 95\% CL)~\cite{OPAL:2002decay},
    Belle $B\to K+{\rm inv.}$ (purple, 90\% CL)~\cite{Belle:2026invisible},
    the Belle II reinterpretation (orange, 95\% CL$_s$)~\cite{Gartner:2026invisible},
    and NA62 $K\to\pi+{\rm inv.}$ (green, 90\% CL)~\cite{NA62:2025hidden}.
    Meson limits assume $\BR(h_2\to{\rm inv.})=1$, with the $B$-decay conversion
    specified in Eq.~\eqref{eq:BKinv_prediction}.
    Blue points satisfy the relic-density and Higgs-invisible requirements discussed in the text.
    Red points A and B are the benchmarks satisfying all constraints considered in this work.
    The curves are displayed only within the available mass intervals.
    }
    \label{fig:mixing}
\end{figure}

For $M_2<M_B-M_K\simeq4.785$ GeV, on-shell $h_2$ production in $B^+\to K^+h_2$
provides an additional constraint when $h_2$ decays invisibly.
The inclusive $B^+\to K^+\nu\bar{\nu}$ branching-fraction measurement
by Belle II~\cite{Belle-II:2023esi} is not used as a fixed upper limit on this two-body decay.
Instead, we use the mass-dependent observed upper limits on
$\BR(B^+\to K^+h_2)\BR(h_2\to{\rm inv.})$ from the dedicated Belle search
(90\% CL)~\cite{Belle:2026invisible} and the Belle II likelihood reinterpretation
(95\% CL$_s$)~\cite{Gartner:2026invisible}.
For the latter we use the narrow-resonance template, represented by $\Gamma_X=0.1$ GeV
at the resolution of that analysis.
The two searches are complementary: the Belle II reinterpretation improves the sensitivity
at low masses and covers the region around $M_2\simeq2$ GeV vetoed in the dedicated search,
whereas Belle gives stronger limits at higher masses.
We convert both $B$-decay limits using the same production approximation as in our scan,
\begin{gather}
\BR(B^+\to K^+h_2)\BR(h_2\to{\rm inv.})
 = C_B(M_2)\sin^2\theta\,\BR(h_2\to{\rm inv.}), \label{eq:BKinv_prediction}\\
C_B(M_2)
 = \frac{0.4\sqrt{[1-(M_K-M_2)^2/M_B^2][1-(M_K+M_2)^2/M_B^2]}}
 {[1-M_2^2/(38\,{\rm GeV}^2)]^2}, \nonumber
\end{gather}
with $M_B=5.279$ GeV and $M_K=0.494$ GeV.
The coefficient $0.4$ follows the small-scalar-mass normalization of
Ref.~\cite{Ovchynnikov:2023scalar}. The mass dependence uses the
Ball--Zwicky single-pole form $f_0(q^2)=f_0(0)/(1-q^2/m_{\rm fit}^2)$,
with $f_0(0)=0.330$~\cite{Ball:2004ye}. 
In Eq.~\eqref{eq:BKinv_prediction},
the fitted pole scale $m_{\rm fit}^2=37.46~\mathrm{GeV}^2$ is approximated
by $38~\mathrm{GeV}^2$.
For the displayed meson limits we take $\BR(h_2\to{\rm inv.})=1$,
appropriate to the predominantly invisible regime studied here,
whereas the model branching ratio is used when evaluating a benchmark prediction.
At lower scalar masses, the NA62 $K^+\to\pi^++{\rm inv.}$ search gives
further 90\% CL limits~\cite{NA62:2025hidden}.
Figure~\ref{fig:mixing} displays these constraints in the $(M_2,\sin^2\theta)$ plane.
The curves retain their individual confidence levels and are shown only over the
mass intervals represented by the digitized published limits, without extrapolation.

\begin{figure}[!t] 
	\centering%
	\includegraphics[width=12cm]{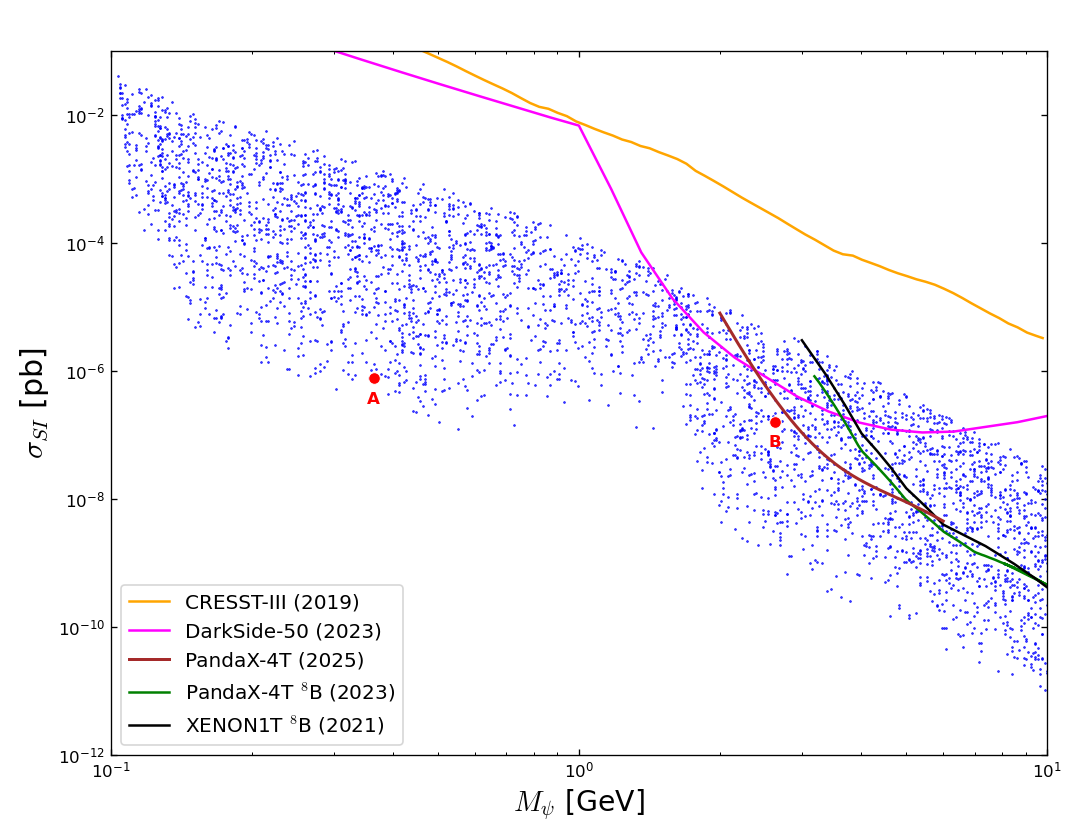}
	\caption{
    Spin-independent DM-nucleon scattering cross section as a function of $M_\psi$.
    Blue points are the same relic-density- and Higgs-invisible-compatible scan points
    as in Fig.~\ref{fig:mixing}.
    The lines show the observed 90\% CL upper limits from
    CRESST-III (2019), DarkSide-50 (2023), PandaX-4T (2025),
    PandaX-4T $^8$B (2023), and XENON1T $^8$B (2021).
    The DarkSide-50 limit includes the Migdal effect and assumes no
    quenching fluctuations (NQ).
    The red points are benchmarks A and B, selected to satisfy all constraints
    considered here and used for the mono-Higgs analysis.
    }
	\label{fig:sigmaSI}
\end{figure}

The precise measurement of the DM relic abundance  
places strong constraints on the allowed parameter space of the model.  
The current value of the dark matter density,  
\(\Omega_{\rm DM} h^2 = 0.1200 \pm 0.0012\),  
has been determined by global fits to cosmological observations,  
primarily based on the Planck satellite data,  
including the temperature and polarization anisotropies of the CMB,  
as well as its gravitational lensing measurements~\cite{Planck:2018vyg}.  
These precise measurements significantly limit the viable range of model parameters  
that can yield the correct thermal relic density.
We note that the DM annihilation cross section in this model is dominated by the $p$-wave contribution,
which is velocity-suppressed at late times.

We calculate the relic abundance with \texttt{micrOMEGAs}~7.1.10
\cite{Belanger:2026MO7}, using a model-level implementation of hadronic
annihilation amplitudes in SFDM. We have adapted the scalar--meson
form factors and data supplied with its \texttt{lightScalar} example
to our two-mediator \texttt{LanHEP}/\texttt{CalcHEP} model
\cite{Semenov:2008jy,Belyaev:2012qa}.
At sub-GeV annihilation energies, free-gluon or light-quark final states
do not reliably describe physical hadrons. We therefore use explicit
$\psi\bar\psi\to h_{1,2}^{*}\to\pi\pi,K\bar K$ amplitudes with
energy-dependent scalar form factors. The implemented channels are
$\pi^+\pi^-$, $\pi^0\pi^0$, $K^+K^-$ and the neutral-kaon pair,
represented by $K_LK_L$ and $K_SK_S$ in the isospin limit,
including the appropriate identical-particle factors.
The reference calculation uses the dispersive form factors of
Winkler~\cite{Winkler:2018qyg}. We compare these with the central BTPZ
and DGL tables supplied with \texttt{lightScalar}, obtained from the
two low-energy matching prescriptions of Blackstone, Tarr\'us
Castell\`a, Passemar and Zupan~\cite{Blackstone:2024hadronic}.
DGL here denotes the Donoghue--Gasser--Leutwyler matching prescription
as implemented in that calculation, rather than leading-order chiral
perturbation theory. The Winkler option additionally includes the
analytic $\eta\eta$ coupling adapted from \texttt{lightScalar},
whereas the BTPZ and DGL options contain the tabulated $\pi\pi$ and $K\bar K$
channels only. This is an exclusive meson-pair treatment, with no
inclusive multihadron contribution added.

For each meson pair, both mediators multiply the same form factor.
Their amplitudes are summed coherently before squaring, retaining
$h_1$--$h_2$ interference. We solve the thermal abundance equation
with \texttt{darkOmegaTR}, starting from the equilibrium abundance.
The meson form factors are evaluated at $\sqrt{s}$ throughout
the thermal average.
We assume standard radiation domination and kinetic equilibrium and
use the package's tabulated effective energy and entropy degrees of
freedom. These thermodynamic inputs are distinct from the hadronic
matrix elements entering the annihilation rate.
We adopt a sharp matching prescription at
$\sqrt{s}_{\rm match}=1.5$~GeV. Below this energy the scalar couplings
to partonic quark and two-gluon final states are switched off and
the meson channels are used. At and above it the meson couplings
are switched off and the model's quark and effective two-gluon
channels are restored. The kinematically accessible leptonic channels
of the model remain active. This mutually exclusive assignment
avoids double counting. The matching scale is an explicit modeling
choice; no smooth matching or matching-scale uncertainty scan is
implied.

For the partonic calculation we retain the effective scalar--gluon
interaction manually implemented in our model, with the usual
heavy-quark effective-operator structure~\cite{Spira:1995rr},
\begin{equation}
\mathcal{L}_{\rm eff}\supset
\frac{1}{4}g_h\bigl(h_1\cos\theta-h_2\sin\theta\bigr)
G^a_{\mu\nu}G^{a\mu\nu},
\label{eq:Leff}
\end{equation}
where $G^a_{\mu\nu}$ is the gluon field strength.
The corresponding SM Higgs coefficient admits the finite-top-mass
expansion,
\begin{equation}
g_h\simeq\frac{\alpha_s}{3\pi v_h}
\left(1+\frac{7}{30}\tau+\frac{2}{21}\tau^2
+\frac{26}{525}\tau^3\right),
\qquad \tau=\frac{M_1^2}{4m_t^2},
\label{eq:gh}
\end{equation}
which approaches the heavy-top limit as $\tau\to0$.
In the numerical model, $g_h$ is treated as a fixed effective coupling
for the two scalar states, with the mixing factors displayed above.
Only the two-gluon vertices of this operator are retained in the
relic-density model. They are multiplied by the same partonic switch
described above, so the explicit $gg$ final state contributes only
for $\sqrt{s}\geq1.5$~GeV. Below this scale the hadronic contribution
is described by the meson form factors.

For benchmark A shown in Fig.~\ref{fig:mixing},
$2M_\psi=0.7285$~GeV and $M_2=0.7316$~GeV.
Near the $h_2$ pole only the pion-pair channels are kinematically
open among the included mesons, and higher-energy configurations are
treated with their actual thresholds in the thermal integral.
Table~\ref{tab:relic_schemes} compares the three prescriptions at
fixed parameters for benchmark A and shows a substantial dependence
on the hadronic form factors. This spread compares central
prescriptions and is not a statistical uncertainty band.
For benchmark B, the reference relic abundance is $\Omega h^2=0.1213$.
Its annihilation threshold $2M_\psi=5.243$~GeV and mediator mass
$M_2=5.581$~GeV both lie above the matching scale, where all three
prescriptions use the same partonic treatment.
This common treatment does not imply an absence of other theoretical
uncertainties.
The scan points and reference benchmarks use the Winkler prescription
and satisfy the Planck $3\sigma$ relic-density constraint,
$0.1164\leq\Omega h^2\leq0.1236$.
The fixed A parameters do not satisfy this interval for BTPZ or DGL,
so its relic-density selection is conditional on the adopted prescription.

\begin{table}[!htb]
\centering
\caption{Relic abundances for benchmark A at fixed model parameters,
with $\sqrt{s}_{\rm match}=1.5$~GeV.
The relative shift is
$\delta_\Omega=100[(\Omega h^2)_{\rm scheme}/(\Omega h^2)_{\rm Winkler}-1]$.
Only central form-factor tables are compared.}
\label{tab:relic_schemes}
\begin{tabular}{lrr}
\hline\hline
Form-factor prescription & $\Omega h^2$ & $\delta_\Omega$ [\%] \\
\hline
Winkler & 0.1173 & 0.00 \\
BTPZ    & 0.06252 & $-46.71$ \\
DGL     & 0.04233 & $-63.91$ \\
\hline\hline
\end{tabular}
\end{table}

The SI cross sections in Fig.~\ref{fig:sigmaSI} show the same scan points as
Fig.~\ref{fig:mixing}, projected onto the $(M_\psi,\sigma_{\rm SI})$ plane.
All blue points satisfy the relic-density and Higgs-invisible requirements.
For comparison, we overlay the 90\% confidence level upper limits  
from CRESST-III (2019) \cite{CRESST:2019jnq}, DarkSide-50 (2023) \cite{DarkSide:2022dhx},  
PandaX-4T (2025) \cite{PandaX:2025light}, PandaX-4T $^8$B (2023) \cite{PandaX:2022aac}, 
and XENON1T $^8$B (2021) \cite{XENON:2020gfr}. 
The scan points displayed in
Figs.~\ref{fig:mixing} and \ref{fig:sigmaSI} lie within the following
parameter ranges:
\beq
\lambda_{2} \in \left[-\pi,\pi\right], \
\lambda_{4} \in \left[0,4\pi\right],  \
M_2/{\rm GeV} \in \left[0.1,25\right], \
M_\psi/{\rm GeV} \in \left[0.1,10\right], \
\tth \in \left[-0.2,~0.2\right]. 
\eeq
The chosen scan range $|\tth|\leq0.2$ is consistent with the OPAL constraint
over the scalar masses considered, but does not represent a
mass-independent experimental upper bound.
With $M_1=125$~GeV and $v_h=246$~GeV, the couplings
$\lambda_0$, $\lambda_1$ and $\lambda_3$ are determined
from the scalar mass matrix as
\begin{equation}
	\begin{aligned}
		\lambda_0 &= \frac{M_1^2\cth^2+M_2^2\sth^2}{2v_h^2},\\
		\lambda_1 &= \frac{(M_1^2-M_2^2)\sth\cth}{v_h}
		-2\lambda_2v_s,\\
		\lambda_3 &= \frac{2}{v_s}
		\left(M_1^2\sth^2+M_2^2\cth^2
		+\frac{\lambda_1v_h^2}{2v_s}
		-\frac{\lambda_4v_s^2}{3}\right).
	\end{aligned}
	\label{eq:derived_couplings}
\end{equation}
We fix $v_s=100$~GeV and $g_S=0.1$ and retain the
points satisfying the relic-density and Higgs-invisible requirements
specified above.
The nonobservation of DM–nucleon scattering events is interpreted as an upper limit on the SI cross sections for a given DM mass.
The DM–nucleon scattering is predominantly mediated 
by a $t$-channel exchange of the $h_2$ boson in most of the phenomenologically allowed parameter space, 
which typically corresponds to the region $M_\psi \approx M_2/2$ 
required to satisfy the relic density constraint.
Figure~\ref{fig:sigmaSI} illustrates the strong impact of direct-detection searches
in the few-GeV part of the scan and their reduced sensitivity at lower DM masses.
For scalar fermion interactions, the leading SI cross section scales as
\begin{equation}
\sigma_{\rm SI}\ \propto\
\mu_{\psi N}^{\,2} g_S^2\sin^2\theta\cos^2\theta
\left(\frac{1}{M_1^2}-\frac{1}{M_2^2}\right)^2,
\label{eq:SI_scaling}
\end{equation}
where $\mu_{\psi N}=M_\psi m_N/(M_\psi+m_N)$ is the DM-nucleon reduced mass.
Thus, for $M_2\ll M_1$, the light-mediator enhancement can produce larger SI
cross sections at lower $M_\psi$ along the near-resonant region $M_2\simeq2M_\psi$.
The weaker experimental sensitivity at low recoil energies is therefore important
even when the predicted scattering cross section is relatively large.
This motivates studying light scalar-mediated fermion DM without introducing a
pseudoscalar interaction to suppress direct detection.
At the same time, on-shell invisible $B$ decays constrain $M_2<M_B-M_K$,
and kaon decays test still lower scalar masses, as shown in Fig.~\ref{fig:mixing}.

The selected range $M_\psi\in[0.1,10]$ GeV, with $M_2\leq25$ GeV and
$2M_\psi<M_2$, also keeps the invisible cascade
$h_1\to h_2h_2\to(\psi\bar\psi)(\psi\bar\psi)$ open and well away from its
$h_1\to h_2h_2$ threshold.
For fixed trilinear coupling $c_{122}$, a smaller $M_2$ increases the phase-space
factor in Eq.~\eqref{eq:h1_h2h2_width}, although this factor is already close to
unity for $M_2\ll M_1/2$.
Unlike the SI scattering amplitude, the dominant on-shell Higgs-pair contribution
to mono-Higgs production is governed mainly by Higgs production and the invisible
Higgs branching ratio, which depends on $c_{122}$ and the scalar masses.
Consequently, the trend in $\sigma_{\rm SI}$ does not imply a universal monotonic
dependence of the mono-Higgs cross section on $M_\psi$ when the other parameters vary.
Our focus on this mass window is motivated by the prospect of sizable
mono-Higgs rates compatible with direct-detection constraints and by
an accessible invisible Higgs cascade.
The flavor limits remain essential to its phenomenology.
The cascade conditions alone require $2M_\psi<M_2<M_1/2$, or
$M_\psi<M_1/4=31.25$ GeV for $M_1=125$ GeV.
The upper scan edge of 10 GeV should not be interpreted as an exclusion limit, 
and no general exclusion of heavier DM is inferred from this analysis.

The red points in Figs.~\ref{fig:mixing} and \ref{fig:sigmaSI} are selected
individually to satisfy all constraints considered in this work.
Benchmark A illustrates the low-mass region in which invisible $B$ decays are
important, whereas benchmark B illustrates the few-GeV DM region with
$2M_\psi>M_B-M_K$, where $B\to K\psi\bar\psi$ is kinematically forbidden
even through an off-shell $h_2$.
For B, $\sigma_{\rm SI}\simeq1.62\times10^{-43}~\mathrm{cm}^2$ at
$M_\psi\simeq2.62$~GeV is approximately $0.44$ times the PandaX-4T (2025)
observed 90\% CL limit shown in Fig.~\ref{fig:sigmaSI}.
Their choice also contrasts two relic-density regimes: A probes
sub-GeV hadronic annihilation sensitive to the form-factor prescription,
whereas B lies above our matching scale and is insensitive to this
choice because the partonic treatment is common to all three prescriptions.
These benchmarks will be used for the mono-Higgs analysis in the next section.

\section{Signatures}\label{sec:signatures}

\begin{figure}[t]
	\centering
	\includegraphics[width=5.8cm]{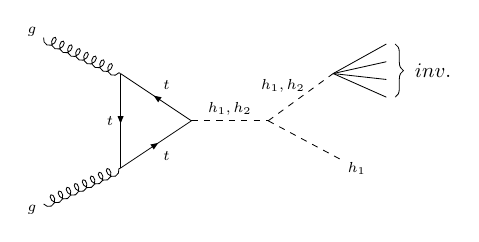} \
	\includegraphics[width=5.8cm]{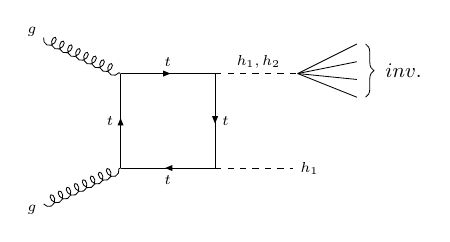} \
	\includegraphics[width=5.8cm]{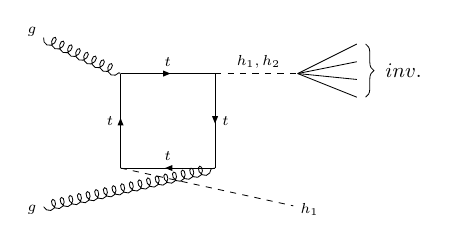}
	\caption{Feynman diagrams for important contributions 
	to mono-Higgs production at the LHC.} 
	\label{fig:feynman}
\end{figure}

In the ATLAS analysis~\cite{ATLAS:2021shl}, the missing transverse momentum
vector $\boldsymbol p_T^{\rm miss}$ is reconstructed from the negative
vector sum of the reconstructed-object momenta and a soft term from
tracks associated with the primary vertex but not with those objects.
Its magnitude is denoted by $E_T^{\rm miss}$.
The mono-Higgs process is characterized by the presence of a Higgs boson 
accompanied by large missing transverse momentum in the final state. 
In our theoretical analysis, $E^{\textrm{miss}}_{\textrm{T}}$ arises 
from the production of pairs of DM fermions through the decays of $h_1$ and $h_2$.  
The dominant sources of mono-Higgs events are the processes 
$pp \to h_1 h_{1,2} \to h_1 \bar{\psi}\psi$ and 
$pp \to h_1 h_1 \to h_1 (h_2 h_2) \to h_1 (\bar{\psi}\psi \bar{\psi}\psi)$.
Our signal definition uses the predominantly invisible $h_2$ regime
specified in Sec.~III.
When $h_2\to\psi\bar\psi$ is closed, its lifetime depends on both its
mass and mixing, and it can be long-lived~\cite{Winkler:2018qyg}.
Visible or displaced $h_2$ decays, and missing momentum from a
detector-stable $h_2$, are outside the scope of this analysis.

The dominant Feynman diagrams are shown in Fig.~\ref{fig:feynman},  
consisting of triangle and box diagrams induced by gluon-gluon fusion.
As in the case of SM di-Higgs production, these contributions exhibit destructive interference.
As a result, the total cross section depends on nontrivial combinations of new model parameters in a complicated manner, 
making it difficult to identify a priori which parameters or regions of parameter space 
play a dominant role in enhancing the mono-Higgs signal.
For the two benchmarks studied below, non-resonant $h_1h_1$ production
dominates the mono-Higgs rate, with the invisible Higgs decay mainly
proceeding through $h_1\to h_2h_2\to(\bar\psi\psi)(\bar\psi\psi)$.
The coupling $c_{122}$ controls this cascade partial width through
Eq.~\eqref{eq:h1_h2h2_width}, while the signal rate also depends on the
production cross sections, the total Higgs width and the direct
invisible decay. We retain both invisible decay modes in the
normalization below.

\begin{figure}[!htb]
    \centering
    \includegraphics[width=12cm]{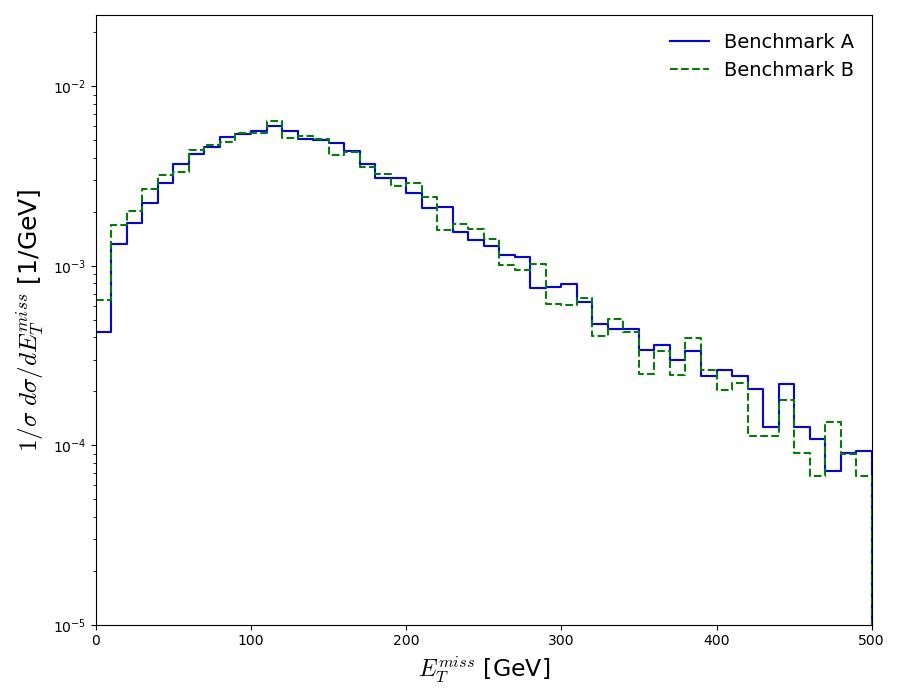}
    \caption{
    Normalized missing-transverse-momentum distributions for inclusive mono-Higgs
    production at $\sqrt{s}=13$~TeV for benchmarks A (blue solid) and B (green dashed).
    The distributions use Delphes samples with an ATLAS-like detector configuration before the
    $b\bar b$ decay requirement and signal-region cuts. Normalization includes
    events outside the displayed range.
    }
    \label{fig:ETmiss0}
\end{figure}

\begin{figure}[!htb]
    \centering
    \includegraphics[width=12cm]{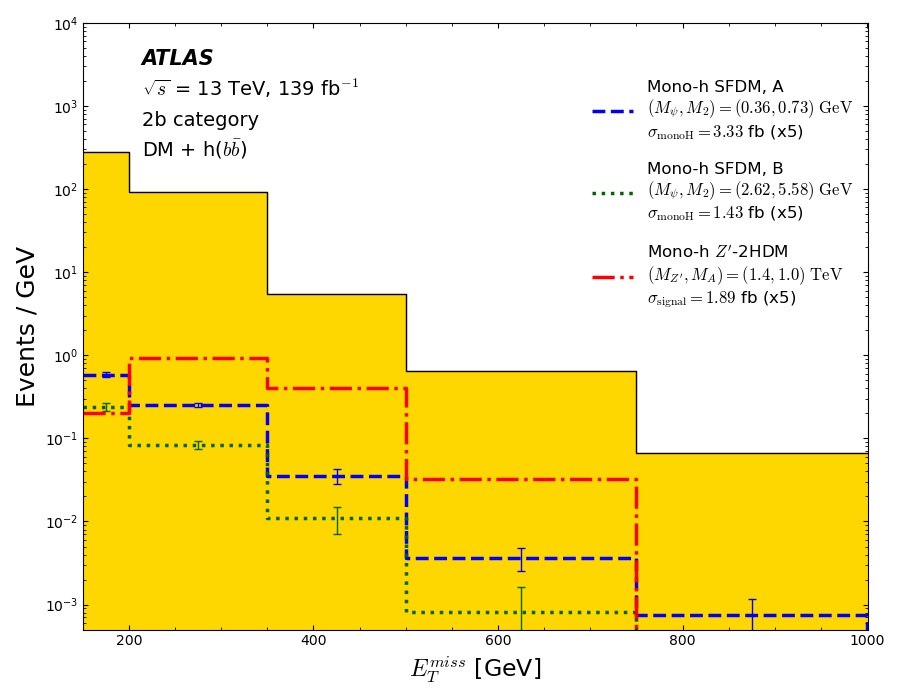}
    \caption{
    Illustrative comparison in the ATLAS $2b$-tag regions at 13~TeV and
    $139~\mathrm{fb}^{-1}$. The SFDM curves for A (blue dashed) and B (green dotted)
    use the partial resolved and merged selections described in the text; error bars
    show MC statistical uncertainties only. The background (gold) is from
    Table~4 of Ref.~\cite{ATLAS:2021shl}, and the ATLAS example $Z'$--2HDM signal
    (red dash-dotted) is taken from Fig.~6(a) of the same reference.
    All signal curves are multiplied by five. The SFDM legend gives
    $\sigma_{\rm monoH}$ before the visible Higgs decay and selection cuts.
    The last A/background bin includes all $E_T^{\rm miss}\geq750$~GeV events,
    displayed with a width of 250~GeV. The digitized reference ends at 750~GeV.
    Vertical closures do not represent additional zero-yield bins.
    The B overflow estimate is statistically unresolved and omitted, as explained in the text.
    }
    \label{fig:ETmiss0e}
\end{figure}

To illustrate this behavior, we analyze the $E_T^{\text{miss}}$ distribution
of mono-Higgs production in the SFDM model using two representative benchmark
points satisfying all constraints discussed in Sec.~\ref{sec:constraints}.
Their parameters are given below, with the relic-density requirement
evaluated using the reference Winkler prescription:
\begin{equation}
\label{eq:bench}
\begin{aligned}
A:\quad & t_\theta=0.001587,\quad M_2=0.7316~{\rm GeV},\quad
M_\psi=0.3643~{\rm GeV},\quad \lambda_2=0.004594,\quad \lambda_4=0.1336;\\[3pt]
B:\quad & t_\theta=0.01614,\quad M_2=5.581~{\rm GeV},\quad
M_\psi=2.621~{\rm GeV},\quad \lambda_2=-0.007344,\quad \lambda_4=3.758.
\end{aligned}
\end{equation}
The common parameters are $M_1=125$~GeV, $v_h=246$~GeV,
$v_s=100$~GeV and $g_S=0.1$.
The benchmark values in Eq.~\eqref{eq:bench} are rounded to four significant
digits, and all numerical results use the unrounded parameters.
Benchmark A probes low-mass DM with $M_2<M_B-M_K$, where rare $B$-meson
decays provide a particularly important constraint on the scalar mixing.
Benchmark B has few-GeV DM with $2M_\psi>M_B-M_K$, 
while direct detection provides a complementary test.
These benchmark points, indicated by red dots in Figs.~\ref{fig:mixing} and \ref{fig:sigmaSI},
correspond to trilinear couplings
$c_{122}=2.260$~GeV for A and $c_{122}=-1.319$~GeV for B,
evaluated from Eq.~\eqref{eq:cijk}.
The Higgs invisible branching fractions are $0.09080$ and
$0.03572$, and the inclusive mono-Higgs cross sections at
$\sqrt{s}=13$~TeV are $3.330$~fb and $1.429$~fb,
respectively, before the visible-Higgs branching fraction and
analysis cuts. 
We use $10^5$ hard events for each of the $h_1h_1$ and $h_1h_2$ processes
at each benchmark, generated with
\texttt{MadGraph5\_aMC@NLO}~3.8.1~\cite{Alwall:2014hca} at $\sqrt{s}=13$~TeV.
We use the built-in NNPDF2.3QED LO parton distributions with
$\alpha_s(M_Z)=0.130$ and dynamical renormalization and factorization scales
$\mu_R=\mu_F=\frac{1}{2}\sum_i\sqrt{m_i^2+p_{T,i}^2}$,
where the sum runs over the two final-state scalars.
The SFDM model is supplied as a UFO implementation generated with
\texttt{FeynRules}~2.3.32~\cite{Alloul:2013bka}, including QCD ultraviolet
counterterms and rational $R_2$ terms. The samples describe loop-induced
leading-order gluon fusion~\cite{Hirschi:2015iia}: the one-loop amplitudes
are the leading contribution in these channels, and NLO QCD corrections
to this contribution are not included.
The triangle and box amplitudes retain the full top- and bottom-quark
mass dependence and their interference, with $m_t=172$~GeV and
$m_b=4.7$~GeV, which also set the corresponding Yukawa inputs.
No heavy-top effective interaction is used for these hard amplitudes.
The effective gluon interaction discussed in Sec.~III pertains to the
partonic relic-density calculation.
No additional $K$ factor is applied to the production cross sections
listed in Table~\ref{tab:point3norm}.
Parton showering and hadronization use
\texttt{Pythia}~8.317~\cite{Sjostrand:2014zea}, and the ATLAS comparison uses
\texttt{Delphes}~3.5.1~\cite{deFavereau:2013fsa} samples with an ATLAS-like detector configuration.

To separate decay sampling from physical normalization, define
$B_1=\Br(h_1\to{\rm invisible})$, including direct and cascade decays,
and $B_2=\Br(h_2\to\psi\bar\psi)$. For production cross sections
$\sigma_{11}=\sigma(pp\to h_1h_1)$ and $\sigma_{12}=\sigma(pp\to h_1h_2)$,
the weights for exactly one visible $h_1$ are
\begin{equation}
 W_{11}=2\sigma_{11}B_1(1-B_1),\qquad
 W_{12}=\sigma_{12}B_2(1-B_1),\qquad
 \sigma_{\rm monoH}=W_{11}+W_{12}.
 \label{eq:point3norm}
\end{equation}
For a visible decay $X$, we approximate its branching fraction conditional
on a non-invisible $h_1$ decay by the SM value,
$q_X=\Br(h_1\to X)/(1-B_1)\simeq\Br_{\rm SM}(h\to X)$.
In particular, $q_{b\bar b}=0.5824$ at $M_1=125.00$~GeV
\cite{LHCHiggsCrossSectionWorkingGroup:2016ypw}.
The cross section and expected count in a reconstructed region $R$ are
\begin{equation}
 \sigma_{X,R}=q_X\sum_{p=11,12}W_p\,\epsilon_{p,R}^{X},
 \qquad
 N_{X,R}=\mathcal{L}\,\sigma_{X,R},
 \label{eq:point3yield}
\end{equation}
where $\epsilon_{p,R}^{X}$ denotes the combined acceptance and selection efficiency
for generated mono-$X$ events from process $p$ to enter region $R$
after the implemented reconstruction and selection.
It is conditional on this decay topology, with all such events, including
those outside the missing-momentum range, in the denominator.
The physical $B_1,B_2$ and $q_X$ are therefore not
replaced by the branching fractions sampled by the shower.
The generated mixture of direct and cascade invisible decays is retained
in these conditional efficiencies.
Table~\ref{tab:point3norm} gives the normalization before analysis cuts.

\begin{table}[!htb]
\centering
\setlength{\tabcolsep}{8pt}
\caption{Physical normalization at 13~TeV before analysis cuts.
$\sigma_{\rm monoH}$ excludes the visible-Higgs branching fraction;
$\sigma_{b\bar b}^{\rm before}=q_{b\bar b}\sigma_{\rm monoH}$ includes it.
The last row gives the unselected count at $139~\mathrm{fb}^{-1}$.}
\label{tab:point3norm}
\begin{tabular}{lrr}
\hline\hline
Quantity & A & B \\
\hline
$B_1$ & $0.09080$ & $0.03572$ \\
$B_2$ & $1$ & $0.99998$ \\
$\sigma_{11}$ [fb] & $20.16$ & $20.15$ \\
$\sigma_{12}$ [fb] & $8.848\times10^{-4}$ & $0.04250$ \\
$\sigma_{\rm monoH}$ [fb] & $3.330$ & $1.429$ \\
$\sigma_{b\bar b}^{\rm before}$ [fb] & $1.939$ & $0.8323$ \\
$N_{b\bar b}^{\rm before}$ & $269.6$ & $115.7$ \\
\hline\hline
\end{tabular}
\end{table}

Figure~\ref{fig:ETmiss0} shows the inclusive distributions before visible-decay
and signal-region requirements. Both spectra peak at
$E_T^{\rm miss}\sim100$--$150$~GeV and fall towards larger momenta.
For ATLAS we implement a partial selection based on Table~1
of Ref.~\cite{ATLAS:2021shl}, separating resolved
($150<E_T^{\rm miss}<500$~GeV) and merged ($E_T^{\rm miss}\geq500$~GeV)
regions. The experimental backgrounds and $Z'$--2HDM reference in
Fig.~\ref{fig:ETmiss0e} use the full selection of that analysis.
In the resolved regions, anti-$k_T$ jets with $R=0.4$ satisfy
$p_T>20$~GeV and $|\eta|<2.5$ (central), or $p_T>30$~GeV and
$2.5\leq|\eta|<4.5$ (forward). We veto reconstructed electrons and muons
with $p_T>7$~GeV in $|\eta|<2.47$ and $2.5$, respectively, and apply
generic hadronic-$\tau$ and track-multiplicity-based extended-$\tau$ vetoes.
The latter rejects a jet within $\Delta\phi<\pi/8$ of the missing momentum
when it contains one to four tracks with $p_T>0.5$~GeV, $|\eta|<2.5$
and $\Delta R<0.4$. For the leading (up to three) small-radius jets we require
$\Delta\phi(\boldsymbol{p}_T^{\rm miss},j)>20^\circ$.
At least two central $b$-tagged jets are required.
The two leading tagged jets form the Higgs candidate.
We require $p_T^h>100$~GeV for $E_T^{\rm miss}<350$~GeV
and $p_T^h>300$~GeV for $E_T^{\rm miss}\geq350$~GeV.
We also require $m_T^{b,\mathrm{near}}>170$~GeV
and $m_T^{b,\mathrm{far}}>200$~GeV, where
$
m_T^b=\sqrt{2p_T^b E_T^{\rm miss}
	\left[1-\cos\Delta\phi(\boldsymbol p_T^b,
	\boldsymbol p_T^{\rm miss})\right]}
$
is the transverse mass formed from a tagged jet and the missing
transverse momentum. The labels near and far refer to the tagged jets
with the smallest and largest azimuthal separations from
$\boldsymbol p_T^{\rm miss}$, respectively.
The central-jet multiplicity is at most four for $2b$
and five for $\geq3b$, with $50<m_h^{\rm reco}<280$~GeV.
Here $m_h^{\rm reco}$ denotes the reconstructed Higgs-candidate mass,
which is $m_{bb}$ in the resolved regions.

In the merged regions, calorimeter towers are clustered into anti-$k_T$
$R=1.0$ jets and trimmed with $k_T$ subjets of $R=0.2$, keeping those with
$p_T^{\rm subjet}/p_T^{\rm parent}>0.05$.
Tracks with $p_T>0.5$~GeV and $|\eta|<2.5$ form variable-radius
anti-$k_T$ track-jets with
$R_{\rm eff}=\min[0.4,\max(0.02,30~{\rm GeV}/p_T)]$,
ghost-associated with the untrimmed parent jet.
Our acceptance assumes a trimmed jet with $p_T>250$~GeV, $|\eta|<2$,
and track-jets with $p_T>10$~GeV, $|\eta|<2.5$ and at least two tracks.
The leading accepted large-radius jet is the Higgs candidate, and its
two leading associated track-jets must be tagged.
The $2b$ category vetoes all other track-jet tags, while $\geq3b$ requires
at least one additional tag outside the Higgs candidate.
We retain the common vetoes and azimuthal requirement and impose
$50<m_h^{\rm reco}<270$~GeV after an approximate muon-in-jet correction.
The resolved transverse-mass and jet-count requirements do not apply here.

The resolved jets use generic Delphes tag/mistag efficiencies.
For merged jets these functions are transferred to track-jets with
ghost-associated heavy-hadron flavour labels; independent tag probabilities
are summed analytically. This is an approximation, not the calibrated
ATLAS DL1 working point. The resolved missing-momentum significance
requirement $\mathcal{S}>12$ is not applied. Trigger efficiencies,
pile-up/jet-vertex tagging, full lepton/$\tau$ identification and overlap
removal, and experiment-specific jet/missing-momentum calibrations are
also not reproduced. No $Z'$--2HDM cutflow validation has been performed.
The comparison is therefore an approximate detector-level study,
not a validated statistical recast.

\begin{table}[!htb]
\centering
\setlength{\tabcolsep}{8pt}
\caption{ATLAS-region comparison at $139~\mathrm{fb}^{-1}$.
Signal cross sections and counts follow our partial selection, with
$50<m_h^{\rm reco}<280$~GeV (resolved) or
$50<m_h^{\rm reco}<270$~GeV (merged),
where $m_h^{\rm reco}$ denotes the reconstructed Higgs-candidate mass.
Signal errors are MC statistical only. Backgrounds and observations are the
full-selection results of Tables~4 and 5 of Ref.~\cite{ATLAS:2021shl}, and
background errors include statistical and systematic uncertainties.
No display factor of five is applied.
\textit{n.r.} marks statistically unresolved signal contributions (see text).}
\label{tab:point3atlas}
\begingroup
\setlength{\tabcolsep}{3pt}
\footnotesize
\begin{tabular}{ccrrrrrr}
\hline\hline
$n_b$ & $E_T^{\rm miss}$ & $\sigma_A$ & $N_A$ & $\sigma_B$ & $N_B$ & Background & Data \\
 & [GeV] & [$10^{-3}$ fb] & & [$10^{-3}$ fb] & & & \\
\hline
$2$ & $[150,200)$ & $42.2\pm3.0$ & $5.86\pm0.41$ & $17.3\pm1.9$ & $2.40\pm0.27$ & $14070\pm110$ & 14259 \\
$2$ & $[200,350)$ & $54.7\pm3.4$ & $7.61\pm0.47$ & $18.1\pm2.0$ & $2.51\pm0.28$ & $13860\pm100$ & 13724 \\
$2$ & $[350,500)$ & $7.69\pm1.60$ & $1.07\pm0.22$ & $2.39\pm0.86$ & $0.33\pm0.12$ & $825\pm19$ & 799 \\
$2$ & $[500,750)$ & $1.310\pm0.408$ & $0.182\pm0.057$ & $0.292\pm0.286$ & $0.041\pm0.040$ & $160\pm8$ & 168 \\
$2$ & $\geq750$ & $0.269\pm0.156$ & $0.037\pm0.022$ & n.r.$^\dagger$ & n.r.$^\dagger$ & $16.7\pm1.9$ & 19 \\
\hline
$\geq3$ & $[150,200)$ & $1.27\pm0.52$ & $0.177\pm0.072$ & $0.226\pm0.222$ & $0.031\pm0.031$ & $430\pm15$ & 408 \\
$\geq3$ & $[200,350)$ & $2.34\pm0.70$ & $0.325\pm0.098$ & $1.17\pm0.52$ & $0.162\pm0.072$ & $656\pm21$ & 658 \\
$\geq3$ & $[350,500)$ & n.r.$^\dagger$ & n.r.$^\dagger$ & n.r.$^\dagger$ & n.r.$^\dagger$ & $42\pm4$ & 42 \\
$\geq3$ & $\geq500$ & $0.017\pm0.011$ & $0.0024\pm0.0015$ & n.r.$^\dagger$ & n.r.$^\dagger$ & $42\pm2.8$ & 46 \\
\hline\hline
\end{tabular}
\endgroup
\end{table}

Table~\ref{tab:point3atlas} gives the cross sections after the implemented
cuts and the signal/background counts in all nine regions.
For each process, we estimate
$\epsilon=\sum_iw_i a_i/\sum_iw_i$ and
$\mathrm{Var}(\epsilon)=\sum_iw_i^2(a_i-\epsilon)^2/(\sum_iw_i)^2$
over all truth-selected mono-$b\bar b$ events,
where $w_i$ is the Monte Carlo weight of event $i$, and $a_i$ is
the resolved selection indicator or the merged category probability
after kinematic cuts, with $a_i=0$ outside the region.
Errors include numerator--denominator correlations but exclude production,
branching-fraction and detector uncertainties. Sparse-bin errors are
large-sample variance estimates. In the dominant $h_1h_1$ component,
A above 750~GeV has about three effective events, while B in the
$500$--$750$~GeV $2b$ bin has only about one. The B resolved $\geq3b$
estimate in the first bin is likewise supported by one $h_1h_1$ event.
Daggered entries have insufficient MC support: no $h_1h_1$ event passes
the resolved $\geq3b$ selection at $350$--$500$~GeV, and for B no
candidate with two leading $b$-flavoured track-jets contributes to the
merged $2b$ overflow or $\geq3b$ category in that dominant process.
Residual mistag and $h_1h_2$ contributions do not constrain the unsampled
$h_1h_1$ double-$b$ contribution. These entries are neither zero-signal
predictions nor physical upper limits.

Figure~\ref{fig:ETmiss0e} displays the $2b$ estimates alongside the ATLAS
reference. The largest counts occur in the resolved bins:
$5.86\pm0.41$, $7.61\pm0.47$, $1.07\pm0.22$ for A and
$2.40\pm0.27$, $2.51\pm0.28$, $0.33\pm0.12$ for B.
The $500$--$750$~GeV estimates are $0.182\pm0.057$ and
$0.041\pm0.040$ events, respectively.
For a rate comparison with the model-independent limits, we also impose
$90<m_h^{\rm reco}<150$~GeV, distinct from the broader mass windows in
Table~\ref{tab:point3atlas}. In the resolved $2b$ bins
$150$--$200$, $200$--$350$ and $350$--$500$~GeV, the estimated cross sections
are $(0.0349\pm0.0027,\,0.0475\pm0.0032,\,0.0064\pm0.0015)$~fb for A and
$(0.0144\pm0.0018,\,0.0156\pm0.0019,\,0.00194\pm0.00080)$~fb for B.
The corresponding observed 95\% CL limits are approximately
$(3.26,\,1.47,\,0.326)$~fb, read from Fig.~9 of
Ref.~\cite{ATLAS:2021shl} with a few-percent digitization uncertainty.
Thus even the largest central signal-to-limit ratio is only $3.2\%$ for A
and $1.1\%$ for B.
In the $500$--$750$~GeV and $\geq750$~GeV $2b$ bins, the estimates for A
are $(1.31\pm0.41)\times10^{-3}$ and $(2.7\pm1.6)\times10^{-4}$~fb,
compared with limits of approximately $0.151$ and $0.050$~fb, respectively.
For B, both narrower merged bins lack double-$b$ support in the dominant
$h_1h_1$ sample and are statistically unresolved.
The different detector selections prevent interpreting these rate comparisons
as a formal exclusion test.

The CMS analysis~\cite{CMS:2019ykj} reconstructs
$h\to ZZ^*\to4e,4\mu,2e2\mu$, including electrons and muons from $\tau$ decays.
Its requirements include two opposite-sign, same-flavour lepton pairs with
$12<m_{\ell\ell}<120$~GeV, $m_{Z_1}>40$~GeV for the pair closest to the
nominal $Z$ mass, leading/subleading lepton $p_T>20/10$~GeV,
$115<m_{4\ell}<135$~GeV,
no additional selected lepton, and fewer than two $b$-tagged jets.
The search uses the selected $p_T^{\rm miss}$ distribution.
We do not simulate this CMS selection. Instead, the small final-state
branching fraction allows an inclusive bound before any acceptance loss.
At $M_1=125.00$~GeV the SM branching fractions are
$q_{4\ell(e,\mu)}=1.240\times10^{-4}$ and
$q_{4\ell(e,\mu,\tau)}=2.745\times10^{-4}$
\cite{LHCHiggsCrossSectionWorkingGroup:2016ypw}.
The first includes direct $4e,4\mu,2e2\mu$ decays, and the second includes all
primary charged-lepton flavours. Counting every primary $\tau$ mode as if
it were accepted as four reconstructed electrons/muons defines the conservative upper bounds
\begin{equation}
 \sigma_{\rm CMS}^{\rm upper}=
 \sigma_{\rm monoH}\,q_{4\ell(e,\mu,\tau)},\qquad
 N_{\rm CMS}^{\rm upper}=
 (35.9~{\rm fb}^{-1})\,\sigma_{\rm monoH}\,q_{4\ell(e,\mu,\tau)}.
 \label{eq:point3cms}
\end{equation}
The superscript ``direct'' denotes the direct $4e$, $4\mu$, and
$2e2\mu$ yields before analysis cuts, while ``upper'' denotes
an upper bound on the CMS signal yield, including primary $\tau$ modes
with unit acceptance and selection efficiency.
The actual yield passing the CMS selection satisfies
$N_{\rm CMS}\leq N_{\rm CMS}^{\rm upper}$.
This bound assumes the SM conditional visible branching fractions and nominal
production rates above. For B we set $B_2=1$, a negligible conservative increase.
The total selected yield is at most $0.0328$ events for A and $0.0141$ for B.
Any individual missing-momentum bin is bounded by this same total, not assigned
that total as its prediction.
Table~\ref{tab:point3cms} contrasts these bounds with the published
full-selection background and observed counts.

\begin{table}[!htb]
\centering
\setlength{\tabcolsep}{8pt}
\caption{CMS four-lepton comparison at $35.9~\mathrm{fb}^{-1}$.
The upper-bound rows give the analytic upper bounds from
Eq.~\eqref{eq:point3cms}, including all primary $e,\mu,\tau$ modes
with unit acceptance and selection efficiency, while the direct
$e,\mu$ rows include only primary $e,\mu$ modes.
Background and data are the full-selection results of Table~7 of
Ref.~\cite{CMS:2019ykj}, summed over the three final states and all
missing-momentum bins.}
\label{tab:point3cms}
\begin{tabular}{lcc}
\hline\hline
Quantity & A & B \\
\hline
$\sigma_{4\ell(e,\mu)}^{\rm direct}$ [fb]
 & $4.129\times10^{-4}$ & $1.772\times10^{-4}$ \\
$N_{4\ell(e,\mu)}^{\rm direct}$ & $0.0148$ & $0.00636$ \\
$\sigma_{\rm CMS}^{\rm upper}$ [fb]
 & $9.140\times10^{-4}$ & $3.923\times10^{-4}$ \\
$N_{\rm CMS}^{\rm upper}$ & $0.0328$ & $0.0141$ \\
CMS background & \multicolumn{2}{c}{$117.8^{+7.5}_{-7.7}$} \\
CMS observed & \multicolumn{2}{c}{$112$} \\
\hline\hline
\end{tabular}
\end{table}

The approximate ATLAS yields and the CMS four-lepton bounds indicate
limited sensitivity to these two benchmarks in the channels considered.
They do not establish a confidence-level exclusion or a generic bound
on $|c_{122}|$. The mono-Higgs rate depends on several scalar-sector
parameters and is dominated here by non-resonant Higgs-pair production
followed by an invisible decay.
As a luminosity-only illustration, we set
$\mathcal L=3~{\rm ab}^{-1}$ while keeping the $13$~TeV cross sections,
branching fractions and implemented selection unchanged.
The resolved $2b$ region with $150<E_T^{\rm miss}<500$~GeV and
$50<m_h^{\rm reco}<280$~GeV then gives approximately $314\pm14$ events for A
and $113\pm9$ for B. These values follow from Eq.~\eqref{eq:point3yield}
and the sum of the first three $2b$ regions in Table~\ref{tab:point3atlas},
while the errors are Monte Carlo statistical errors on that sum.
The inclusive four-lepton upper bounds from Eq.~\eqref{eq:point3cms},
with unit acceptance for all primary $e,\mu,\tau$ modes, similarly
become approximately $2.74$ and $1.18$ events for A and B.
Increasing the luminosity does not improve the relative precision of
the existing Monte Carlo samples.
These numbers are event-count illustrations at fixed $13$~TeV,
not a $14$~TeV HL-LHC sensitivity projection.
The published backgrounds in Tables~\ref{tab:point3atlas}
and~\ref{tab:point3cms} use the full experimental selections,
whereas our ATLAS signal selection is partial and the CMS signal is
an inclusive upper bound. They therefore do not provide matched
signal and background inputs for a significance calculation.
A quantitative HL-LHC reach requires $14$~TeV signal and background
predictions with the same validated selection, detector and pile-up
response, and a model of background systematic uncertainties and their
correlations. These inputs are not available in the present analysis,
so we do not assign a discovery or exclusion significance.

The benchmarks provide concrete targets for Higgs-pair
production with an invisible decay, as studied in Ref.~\cite{Banerjee:2016nzb}.
The $h_1h_2$ channel contributes only $0.02\%$ (A) and $2.9\%$ (B) of
$\sigma_{\rm monoH}$, so Eq.~\eqref{eq:point3norm} is well approximated by
$\sigma_{\rm monoH}\simeq2\sigma_{11}B_1(1-B_1)$, with
$\sigma_{11}\simeq20.2$~fb for both benchmarks.
The full rates in Table~\ref{tab:point3norm}, $3.33$ and $1.43$~fb,
correspond to invisible branching fractions of $9.1\%$ and $3.6\%$.
This links the invisible-Higgs branching fraction to a quantitative
mono-Higgs signal for light-DM benchmarks selected using the reference
relic-density prescription and the constraints of Sec.~III.
The spectra in Fig.~\ref{fig:ETmiss0} peak at
$E_T^{\rm miss}\sim100$--$150$~GeV, and approximately $93\%$ (A) and
$94\%$ (B) of our estimated resolved $2b$ yield in
Table~\ref{tab:point3atlas} lies in $150<E_T^{\rm miss}<350$~GeV.
This concentration motivates dedicated studies of selections exploiting
the Higgs-pair topology at moderate missing momentum. The benchmark rates
and distributions
provide concrete inputs for such search optimization.
Benchmark A could also be tested through single-Higgs invisible-decay
searches at the HL-LHC. Its $B_1\simeq9.1\%$ exceeds the projected
95\% CL upper limit of $3.8\%$ from the CMS vector-boson-fusion study
at $14$~TeV with $3~\mathrm{ab}^{-1}$, assuming SM Higgs
production~\cite{CMS:FTR18016}.
These searches probe the same invisible branching fraction that governs
the dominant mono-Higgs contribution studied here.

\section{Concluding Remarks}

We have studied mono-Higgs production in the SFDM model at the LHC. 
In this model, a SM gauge-singlet Dirac fermion serves as the DM candidate, 
and its interactions with the visible sector are mediated by a real singlet scalar 
that mixes with the SM Higgs boson. 
This setup not only alters Higgs phenomenology 
but is also subject to various experimental constraints, 
including limits from Higgs decay properties, invisible decay bounds, rare $B$-meson decay constraints,
relic density measurements, and direct detection experiments.

Focusing on light DM, we have studied the mono-Higgs signature from
on-shell $h_1h_1$ and $h_1h_2$ production at $\sqrt{s}=13$~TeV,
followed by direct or cascade decays to DM.
Two representative benchmarks satisfying the relic-density, Higgs, scalar-mixing,
flavor and direct-detection constraints considered were used for the collider study.
We present their inclusive missing-momentum spectra, physical branching fractions,
cross sections before and after the implemented ATLAS cuts, and region-by-region
signal counts alongside the published backgrounds and observations.
The resolved and merged selections are explicitly identified as partial detector
approximations. In the CMS four-lepton channel the inclusive, unit-acceptance
bounds are only $0.0328$ and $0.0141$ events for A and B at
$35.9~\mathrm{fb}^{-1}$. These results indicate limited sensitivity to the
benchmarks in the searches considered, without constituting a formal exclusion
test or a model-wide bound.

\begingroup\clubpenalty=10000
In Higgs-like mediator models, scalar mediators couple preferentially to heavy quarks, 
motivating searches for DM produced in association with a single top quark or a top-quark pair at the LHC
\cite{CMS:2025ncs,CMS:2025izi,ATLAS:2022znu,ATLAS:2022ygn,ATLAS:2020yzc,CMS:2019zzl,CMS:2018ysw}.
These searches provide complementary probes of scalar-mediated DM.
Our study focuses on the light-mediator benchmarks in Eq.~\eqref{eq:bench},
where the relic abundance favors $M_2\simeq2M_\psi$ and the
$h_1\to h_2h_2$ cascade is open.
The luminosity-only illustration in Sec.~IV does not establish an
HL-LHC discovery or exclusion reach. Such a conclusion requires a
dedicated $14$~TeV analysis with validated signal and background
selections, detector and pile-up effects, and background systematic
uncertainties.
The main outcome is a pair of light-SFDM benchmarks connecting the
reference thermal-relic calculation and the constraints of Sec.~III to
inclusive mono-Higgs rates of $1.43$--$3.33$~fb at $13$~TeV with the present
loop-induced leading-order normalization, correlated with invisible-Higgs
branching fractions of $3.6$--$9.1\%$.
These quantitative targets and their moderate-missing-momentum spectra
provide a basis for searches adapted to Higgs-pair production with
invisible decays.

\par\endgroup

\def\npb#1#2#3 {Nucl. Phys. B {\bf#1}, #2 (#3)}
\def\plb#1#2#3 {Phys. Lett. B {\bf#1}, #2 (#3)}
\def\prd#1#2#3 {Phys. Rev. D {\bf#1}, #2 (#3)}
\def\jhep#1#2#3 {J. High Energy Phys. {\bf#1}, #2 (#3)}
\def\jpg#1#2#3 {J. Phys. G {\bf#1}, #2 (#3)}
\def\epj#1#2#3 {Eur. Phys. J. C {\bf#1}, #2 (#3)}
\def\arnps#1#2#3 {Ann. Rev. Nucl. Part. Sci. {\bf#1}, #2 (#3)}
\def\ibid#1#2#3 {{\it ibid.} {\bf#1}, #2 (#3)}
\def\none#1#2#3 {{\bf#1}, #2 (#3)}
\def\mpla#1#2#3 {Mod. Phys. Lett. A {\bf#1}, #2 (#3)}
\def\pr#1#2#3 {Phys. Rep. {\bf#1}, #2 (#3)}
\def\prl#1#2#3 {Phys. Rev. Lett. {\bf#1}, #2 (#3)}
\def\ptp#1#2#3 {Prog. Theor. Phys. {\bf#1}, #2 (#3)}
\def\rmp#1#2#3 {Rev. Mod. Phys. {\bf#1}, #2 (#3)}
\def\zpc#1#2#3 {Z. Phys. C {\bf#1}, #2 (#3)}
\def\cpc#1#2#3 {Chin. Phys. C {\bf#1}, #2 (#3)}
\def\jcap#1#2#3 {JCAP {\bf#1}, #2 (#3)}
\def\compc#1#2#3 {Comput. Phys. Commun. {\bf#1}, #2 (#3)}
\def\pdu#1#2#3 {Phys. Dark Univ. {\bf#1}, #2 (#3)}

\acknowledgments
We acknowledge the use of GPT-6 (OpenAI) and Claude Opus 5.5 (Anthropic) to assist with the data analysis in this work. 
All AI-assisted analyses and numerical results were independently checked and verified by the authors, 
who take full responsibility for the scientific content presented here. 
This work was partially supported by access to OpenAI models provided through the ChatGPT for Academic Researchers program, 
and by Basic Science Research Program through the National Research Foundation of Korea (NRF) 
funded by the Ministry of Education under the Grant No. RS-2023-00248860 (S.-h.N.)
and also funded by the Ministry of Science and ICT under the Grants No. RS-2021-NR059413 (K.Y.L.).

\end{document}